\documentstyle[eqsecnum,prd,aps]{revtex}
\input epsf
\begin{document}
\draft
\title{String Evolution in Open Universes}
\author{C. J. A. P. Martins\thanks{Also at C. A. U. P.,
Rua do Campo Alegre 823, 4150 Porto, Portugal.
Electronic address: C.J.A.P.Martins\,@\,damtp.cam.ac.uk}}
\address{Department of Applied Mathematics and Theoretical Physics\\
University of Cambridge\\
Silver Street, Cambridge CB3 9EW, U.K.}
\maketitle

\begin{abstract}
The velocity-dependent `one-scale' model of Martins \& Shellard is
used
to study the evolution of a cosmic string network (and the
corresponding
loop population) in open universes. It is shown that in this case
there
is no linear scaling regime and that even though curvature still
dominates
the dynamics, at late times strings become the main component of the
universe. We also comment on the possible consequences of these
results.
\end{abstract}

\section{Introduction}
\label{s-int}
Despite the strong theoretical prejudices favouring a flat
universe\cite{kt,li},
there is a fair amount of observational data which suggests
the possibility of an open universe, with a present density that
could be
as low as $\Omega\sim0.3$---notably, the so-called `age
problem'\cite{age} and recent
measurements of the baryonic content of x-ray clusters\cite{xray}.
Since in a low-$\Omega$ universe structures collapse earlier,
observations of
galaxies at high redshift would also be easier to explain if the
universe is
open. It is therefore appropriate to consider how
some of the standard cosmological scenarios would change if this
possibility turns out to be true.

One relevant case is that of the evolution of a network
of cosmic strings\cite{vs}.
It has been shown\cite{def,def1} that defect models
normalized to COBE in an open universe predict a galaxy power
spectrum consistent with that inferred from galaxy surveys without
requiring an extreme bias (in general, $\Omega=1$ models
predict more small-scale power than low-$\Omega h$ ones). However,
these results were established either using {\em a priori}
scaling assumptions for the string network\cite{def} or numerical
simulations of texture evolution\cite{def1}.
Here we study the evolution of a cosmic string network
in open universes, using the velocity-dependent `one-scale'
model of Martins \& Shellard\cite{ms,ms1}, which provides the
first quantitative description of the complete evolution of
the large-scale properties of a cosmic string network.
This is briefly summarised below, and used in the following section
to obtain the evolutionary properties of both the
long-string and the loop populations in an open universe;
these are then compared with the standard flat universe case,
and some implications of these results are discussed.

\section{The velocity-dependent `one-scale' model}
\label{s-mod}
Two different but complementary approaches have been used to study
cosmic
string evolution. The simplest (although more expensive) is to use
large numerical simulations\cite{bb}. Among other
interesting things, these
revealed a significant amount of small-scale structure (or `wiggles')
on
the strings, containing up to one half of the total string energy.

On the other hand, there is always thew possibility of using analytic
methods---an approach first used by Kibble\cite{k85}.
Due to the strings' statistical nature, what one really does is
`string
thermodynamics', that is describing the network by a small number
of macroscopic (or `averaged') quantities whose evolution equations
are
derived from the microscopic string equations of motion, and
introducing
additional `phenomenological parameters' if necessary. The first such
model providing a quantitative picture of the
complete evolution of a string network (and the corresponding loop
population) has been recently developed by Martins \&
Shellard\cite{ms,ms1};
this has the added advantage of being equally applicable to the
study of vortex-string evolution in a condensed matter
context\cite{ms2}.
Here we will present a very brief description of this model---the
reader is
referred to the original paper\cite{ms1} for further details.

Apart from the straightforward definition of the energy of a piece of
string, $E=\mu a(\tau)\int\epsilon d\sigma$
($\epsilon$ being the coordinate energy per unit $\sigma$), the only
other macroscopic quantity in this model is the string RMS velocity,
defined by
\begin{equation}
v^2=\frac{\int{\dot{\bf x}}^2\epsilon d\sigma}{\int\epsilon d\sigma}
\, . \label{vv}
\end{equation}

Explicitly distinguishing between long (or `infinite') strings
and loops, we can use the fact that the former should be Brownian
to define the long-string correlation length as
$\rho_{\infty}\equiv\mu/L^2$.
Phenomenological terms must be included for the interchange
of energy between long strings and loops. A `loop chopping
efficiency' parameter
(expected to be slightly smaller than unity) is introduced to
characterise
loop production
\begin{equation}
\left(\frac{d\rho_{\infty}}{dt}\right)_{\rm to\ loops}=
{\tilde c}v_\infty\frac{\rho_{\infty}}{L}
\, , \label{rtl}
\end{equation}
and in the particular case of GUT-scale strings (but not more
generally\cite{ms1})
it can be safely assumed that loop reconnections onto the long-string
network are negligible. In particular, this has been confirmed in
numerical
simulations\cite{bb}. Note that it is conceivable that the
behaviour of ${\tilde c}$ is different in flat and open universes.
However, this effect will not be crucial, because the
scaling properties do not depend strongly on it\cite{ms1}.

It is then simple to derive the evolution equation for the
correlation
length $L$. Since we are only interested in the epoch around
radiation-matter
equality, we need not be considering frictional forces\cite{ms1}, and
we simply have
\begin{equation}
2 \frac{dL}{dt}=2HL(1+v_\infty^2)+{\tilde c}v_\infty \, . \label{evl}
\end{equation}

For the case of string loops, the relevant lengthscale
is simply the loop length, which decays due to gravitational
radiation,
and its evolution equation is
\begin{equation}
\frac{d\ell}{dt}=(1-2v^2_\ell)H\ell -\Gamma'G\mu v^6_\ell \, ,
\label{evlps}
\end{equation}
where $\Gamma'\sim8\times65$. Then the only other thing that is
needed is
an assumption on the loop
size at formation. In the epoch relevant to this paper, we expect it
to be
approximately constant and much smaller than the correlation
length---we
will take $\ell_i=10^{-3}L(t_i)$. Then for any given time, one only
has to
look at the loops
that have formed until then, determine which of them are still around
and
add up their lengths to determine the total energy density in the
form of loops.
This is conveniently expressible in terms of the ratio of the energy
densities in loops and long strings
\begin{equation}
\varrho(t)\equiv\frac{\rho_o(t)}{\rho_{\infty}(t)}=
g{\tilde c}L^2(t)\int_{t_c}^{t}\frac{a^3(t')}{a^3(t)}
\frac{v_\infty (t')}{L^4(t')}\frac{\ell(t,t')}{\alpha (t')}dt'
 \, , \label{ratio}
\end{equation}
where $g$ is a Lorentz factor accounting for the fact that loops are
usually
produced with a non-zero centre-of-mass velocity, $\alpha$ is the
loop
size at formation relative to the correlation length at that time and
$\ell(t,t')$ is the length at time $t$ of a loop that was formed at a
time $t'$.

Finally, one can derive an evolution equation for the long string or
loop
velocity with only a little more than Newton's second law
\begin{equation}
\frac{dv}{dt}=\left(1-v^2\right)\left(\frac{k}{R}-2Hv\right) \, ;
\label{evv}
\end{equation}
here $k$ is another phenomenological parameter that is related to the
presence
of small-scale structure on the strings; an appropriate ansatz for it
is (refer to \cite{ms1} for a complete justification)
\begin{equation}
k=\left\{ \begin{array}{ll}
1 \, ,& \mbox{$2HR>\chi$} \\
\frac{1}{\sqrt2}2HR \, ,& \mbox{$2HR<\chi$}
 \end{array} \right.
\label{fkans}
\end{equation}
where $R$ is the curvature radius of the string (that is, $R=L$ for
long
strings, but $\ell=2\pi R$ for loops) and $\chi$ is a numerically
determined
coefficient of order unity, whose precise value depends on whether
one is using
the above ansatz for long strings or loops---see\cite{ms1} for a
complete
discussion of this point.

The above quantities are sufficient to quantitatively
describe the large-scale characteristics of a cosmic string network
around the epoch of equal matter and radiation densities (see\cite{ms}).
In a more general situation one would need to include the
effect of frictional forces\cite{ms1} due to particle scattering on
the strings.

\section{Evolution in an open universe}
\label{s-op}
We now come to the issue of this paper. As we already
pointed out, we only need to study the behaviour of the string
network in the transition between the radiation- and the
matter-dominated
regimes (see Martins \& Shellard\cite{ms1}
for a detailed discussion of the early stages of evolution of
GUT-scale and other cosmic string networks).

\begin{figure}
\vbox{\centerline{
\epsfxsize=0.5\hsize\epsfbox{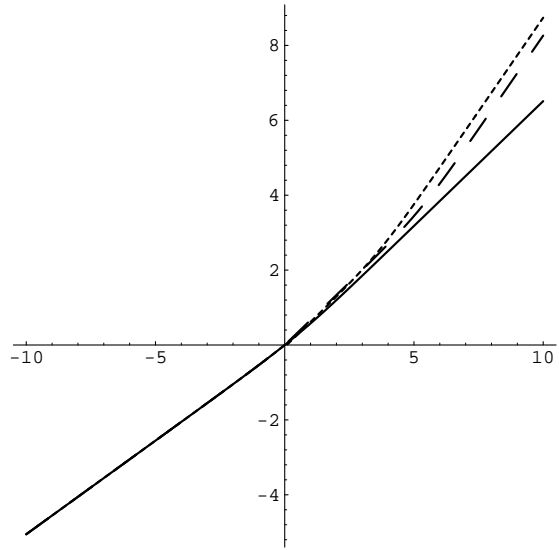}}
\vskip.4in}
\caption{Log-log plot of the scale factor $a$
(relative to $a(t_{eq})=1$) as a function of cosmic time
$t$ (relative to $t_{eq}$) for
a flat universe (solid line) and open universes with a present
density $\Omega_o=0.3$ (dashed line) and $\Omega_o=0.1$ (dotted
line);
both of these have $h=0.6$\,.}
\label{figopenu}
\end{figure}

It is straightforward to see that there is one crucial
difference with respect to the case of a flat universe:
in an open universe there
will no longer be a linear scaling regime. This arises naturally from
the fact that in a universe where the scale factor grows as
$a\propto t^\lambda$ (with $\lambda<1$) the linear regime  has the
following properties
\begin{equation}
\frac{L}{t}=\left[\frac{k(k+{\tilde c})}{4\lambda(1-\lambda)}
\right]^{1/2}  \, ,
\label{openl}
\end{equation}
\begin{equation}
v=\left[\frac{k(1-\lambda)}{\lambda(k+{\tilde c})}
\right]^{1/2}  \, .
\label{openv}
\end{equation}
In an open universe the
`effective' $\lambda$ is a variable, increasing from $\lambda=1/2$ in
the radiation era to an asymptotic value of $\lambda=1$ (see figure
\ref{figopenu}). For example, if we happen to live in a universe
where
$\Omega_o\sim0.3$, we have $\lambda_o\sim0.8$ today, and
$\lambda_o\sim0.9$
for $\Omega_o\sim0.1$.
In other words, there will be corrections to the simple linear
behaviour, such that the
correlation length
$L$ will grow slightly faster than $t$. Nevertheless, since the
horizon size
for an $a\propto t^\lambda$ universe is
\begin{equation}
d_H=\frac{t}{1-\lambda}  \, ,
\label{opendh}
\end{equation}
one can easily show that $L$ will always be smaller than the horizon.
On the other hand, the string velocity will
decrease with time, and therefore loop formation will gradually
switch off.

Note that the power-law dependence of the scale factor obviously
changes
in the transition between radiation and matter domination, even in
the case of
a flat universe. In particular, there will of course be a departure
from linear
scaling while the transition is taking place, with $L/t$ growing from
$0.27$ to $0.6$ (approximately).

In particular, we can easily find the solution to the averaged
equations
of motion in the limit where $\lambda=1$,
\begin{equation}
L\propto t\,\left(\ln t\right)^{1/2}  ,\, \qquad v\propto
\left(\ln t\right)^{-1/2}\, ;
\label{lameq1}
\end{equation}
note that in this limit the correlation length grows more slowly than
the horizon (which goes like $d_H\propto t\ln t$).

In order to get quantitative results, we must solve the averaged
evolution equations described in Section \ref{s-mod} numerically.
To these we must add a further equation---the
Friedmann equation---specifying
how the scale factor (and hence the Hubble parameter) evolves in the
transition between the radiation- and the matter-dominated epochs:
\begin{equation}
H^2+\frac{K}{a^2}=\frac{8\pi G}{3}\left(\rho_{rad}+\rho_{mat}+
\rho_{string}\right) \, , \label{friedm}
\end{equation}
where $K=0$ for a flat universe and $K=-1$ for an open universe;
note that we should at least consider the possibility of the string
density
becoming a non-negligible source for the Friedmann equation.

\begin{figure}
\vbox{\centerline{%
\hskip4em\epsfxsize=.6\hsize\epsfbox{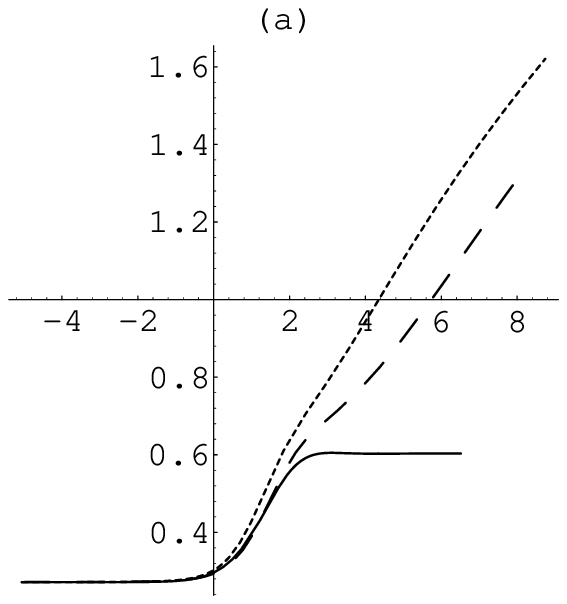}
\hskip-10em\epsfxsize=.6\hsize\epsfbox{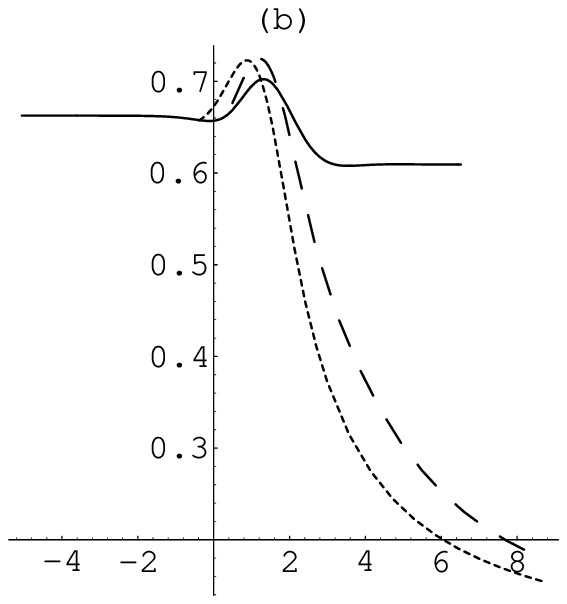}}
\vskip-1in}
\vbox{\centerline{
\hskip4em\epsfxsize=.6\hsize\epsfbox{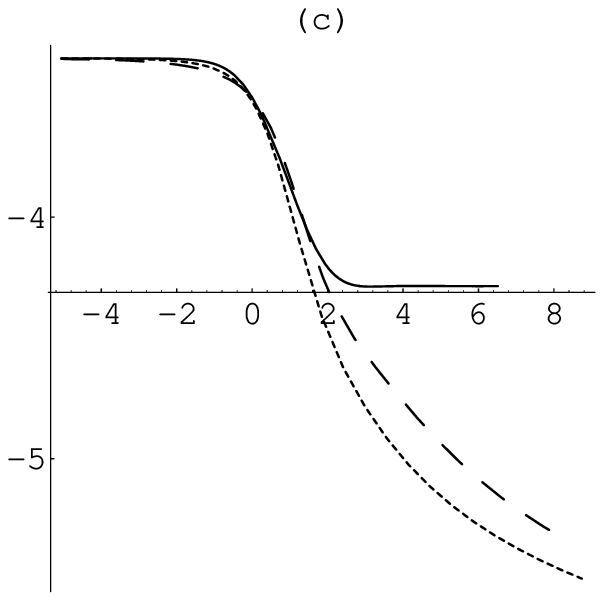}
\hskip-10em\epsfxsize=.6\hsize\epsfbox{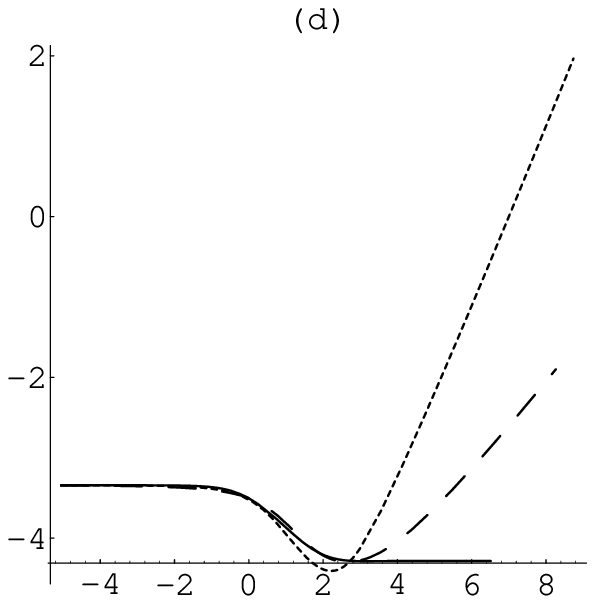}}
\vskip-.75in}
\caption{Properties of a GUT long-string network in a flat universe
(solid)
and in open universes with a present density $\Omega_o=0.3$ (dashed)
and
$\Omega_o=0.1$ (dotted), with $h=0.6$ in both cases.
Plots represent the ratio $L/t$ (a), the RMS string velocity (b), and
the log of the ratio of the long-string density to the
critical (c) and the background (d) densities. The horizontal
axis is labelled in terms of the logarithm of the scale factor
(with $a(t_{eq})=1$); all plotted curves span the period between
$10^{-10}t_{eq}$ and $10^{10}t_{eq}$.}
\label{openlong}
\end{figure}

\begin{figure}
\vbox{\centerline{%
\epsfxsize=.6\hsize\epsfbox{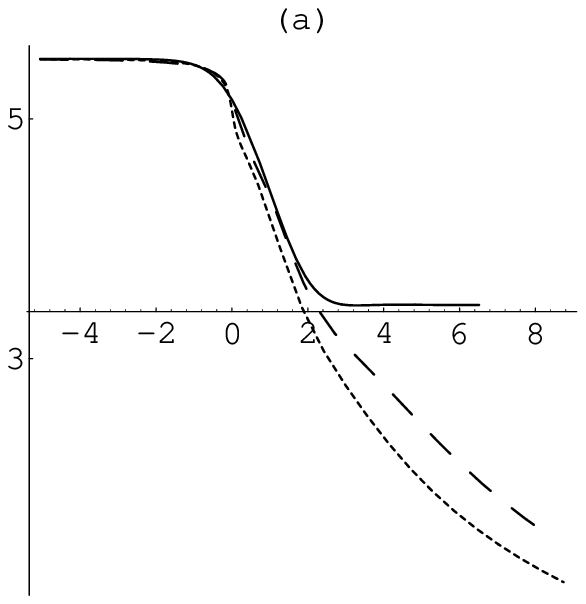}}
\vskip-1in}
\vbox{\centerline{%
\hskip4em\epsfxsize=.6\hsize\epsfbox{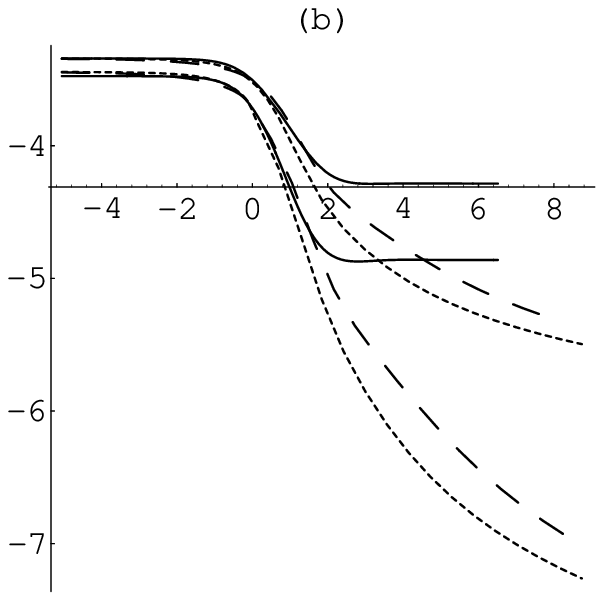}
\hskip-10em\epsfxsize=.6\hsize\epsfbox{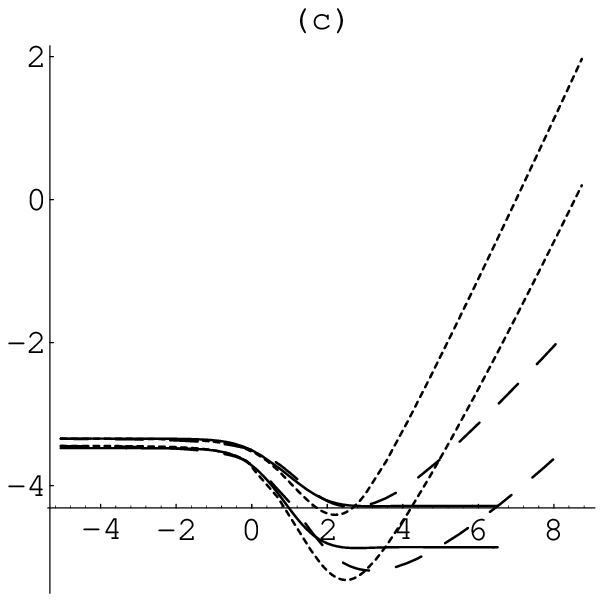}}
\vskip-.75in}
\caption{Properties of the loop population of a GUT string network
in a flat universe (solid) and in open universes with a present
density
$\Omega_o=0.3$ (dashed) and $\Omega_o=0.1$ (dotted), both having
$h=0.6$.
Plot (a) depicts the log of the number of loops produced per Hubble
volume per
Hubble time, while (b)
shows the log of the ratio of the long-string (upper curve of each
pair) and
loop (lower curve of each pair) densities to the critical (b)
and the background (c) densities. The horizontal
axis is labelled in terms of the logarithm of the scale factor
(with $a(t_{eq})=1$); all plotted curves span the period between
$10^{-10}t_{eq}$ and $10^{10}t_{eq}$.}
\label{openloop}
\end{figure}

Figures \ref{openlong} and \ref{openloop} contrast the evolution of
the
long-string and loop populations in a flat universe and in open
universes with $\Omega_o=0.3$ and $\Omega_o=0.1$ (we have assumed
that $h=0.6$
in both cases); note that the present epoch corresponds to
$a_o/a_{teq}\sim2.3\times10^4\, \Omega_o h^2 $.

As was first discussed in\cite{ms1}, even in the case of a flat
universe
the transition from the radiation to the matter epoch is a very slow
process,
lasting about eight orders of magnitude in time. In the case of an
open
universe, apart from the differences we already expected,
the most interesting
result is that, although the string density always decreases with
respect
to the critical density (as one would also expect),
at a redshift around $z\sim\Omega_o^{-1}$  (which is approximately
when curvature
has started to dominate the dynamics) the string density has started
to
grow relative to the background density. In fact, in a $\Omega_o=0.3$
universe strings will become the main component of the
universe in about seven orders of magnitude
in time, whereas if we had $\Omega_o=0.1$ this would only take about
four
orders of magnitude in time.

It is also interesting to point out that, although having
$\Omega_o=0.3$
or $\Omega_o=0.1$ does not yield very significant differences in the
values
of the long-string correlation length or velocity, it does produce
very significant differences in the ratio of the long-string and loop
densities to that of the background.

Moreover, one should note that despite the significant drop in the
number of
loops produced (see \ref{openloop}(a)), the ratio of the energy
densities in
loops and long strings decreases rather more slowly. This is because
loops are slightly larger at formation and, since the average
long-string velocity is decreasing, so is that of large enough
loops (note that we assume that the RMS loop velocity at formation is
equal
to the RMS long string velocity at that time). Consequently, loops
will
live longer, since the redshift and gravitational radiation terms
in (\ref{evlps}) are velocity-dependent.

\section{Conclusions}
\label{s-cc}
In this paper we presented the first discussion of cosmic string
evolution in open universes, in the context of the
generalized `one-scale' model of Martins \& Shellard\cite{ms1}.
We have shown that there is no linear
scaling regime in an open universe, and that although the string
density
always decreases with respect to the critical density, it has been
increasing
relative to that of the background from $z\sim\Omega_o^{-1}$, and it
will
become the main component of the universe sometime in the future.

These differences with respect to the
standard (flat universe) case only become significant fairly late
in the matter-dominated epoch, so with respect to the string-seeded
structure
formation scenario we should only expect changes on very large
scales---that
one can easily estimate to be larger than the scales of the largest
existing
surveys.

On the other hand, such large scales are of course relevant when one
is
comparing the cosmic microwave background anisotropies produced by
cosmic strings to COBE data---thereby normalizing the string mass per
unit
length\cite{acssv}---since this essentially involves an integration
from the
present time to the surface of last scattering. Thus the changes in
the string
network properties discussed in the present paper can significantly
alter this
normalization. This issue will be discussed in a forthcoming
publication.

\acknowledgments
The author thanks Paul Shellard and Pedro Avelino for many
enlightening
discussions. This work was funded by JNICT (Portugal) under `Programa
PRAXIS XXI' (grant no. PRAXIS XXI/BD/3321/94).

\vfill\eject
\vfill\eject

\end{document}